
\input harvmac
\noblackbox
\def\efi{\vbox{\sl\centerline{Enrico Fermi Institute,University
 of Chicago}%
\centerline{5640 Ellis Avenue,Chicago, Illinois 60637}}}

\lref\uno{J.A. Harvey and A. Strominger, Commun. Math.
 Phys. 151(1993) 221.}
\lref\duo{J.P. Gauntlett, {\it {Low Energy Dynamics of
Supersymmetric Solitons}}, preprint EFI-92-25.}
\lref\thr{G.W. Gibbons and N. Manton, Nucl. Phys. B274 (1986) 183.}
\lref\fou{J.P. Gauntlett, {\it {Low Energy Dynamics of N=2
Supersymmetric Monopoles}}, preprint EFI-93-09.}
\lref\fiv{H. Osborn, Phys. Lett. 83B (1979) 321.}
\lref\sox{L. Brink, J.H. Schwarz and J. Scherk,
Nucl. Phys. B121 (1977) 77\semi F. Gliozzi, D. Olive and
J. Scherk, Nucl. Phys. B122 (1977) 253.}
\lref\sev{E. B. Bogolmol'nyi, Sov.J. Nucl. Phys. 24 (1976) 449
\semi M.K. Prasad and C. M. Sommerfield, Phys. Rev. Lett. 35 (1975)
760 \semi G. 'tHooft, Nucl. Phys. B79 (1979) 274 \semi
A. M. Polyakov, JETP Lett. 20 (1974) 194.}
\lref\eig{R. Rajaraman, {\it {Solitons and Instantons}},
North-Holland 1982.}
\lref\nin{M. F. Atiyah and N. Hitchin, {\it {The Geometry and
Dynamics of Magnetic Monopoles}}, Princeton University Press
(1988).}
\lref\dix{C. Callias, Commun. Math. Phys. 62 (1978) 213.}
\lref\ele{C. Montonen and D. Olive, Phys. Lett. 72B (1977) 117.}

\Title{\vbox{\baselineskip12pt
\hbox{EFI-94-04} \hbox{hep-th/9401133} }}
{\vbox{\centerline{Supersymmetric Quantum Mechanics}
\centerline{of Monopoles}
\centerline{in N=4 Yang-Mills Theory}}}

{\bigskip
\centerline{Julie D. Blum}
\bigskip
\efi

\bigskip
\medskip
\centerline{\bf Abstract}

A supersymmetric collective coordinate expansion of
the monopole solution of $N=4$ Yang-Mills theory is
performed resulting in an $N=4$ supersymmetric quantum mechanics on
the moduli space of monopole solutions.}

\Date{1/94}

\newsec{Introduction}

This article is an extension of the work of Harvey,
Strominger \uno, and Gauntlett \duo determining the low energy
Lagrangians for solitons which are solutions of supersymmetric
theories.  Here we focus on monopoles in the $N=4$
supersymmetric Yang-Mills theory.  What we find is an $N=4$
 supersymmetric quantum mechanics of monopoles dancing
on their moduli space.  The $N=0$ (bosonic) and $N=2$
cases have been previously handled by Gibbons and Manton \thr and
Gauntlett \fou .

In this low energy approximation one considers only the dynamics
 of monopole zero modes, fields with classical trajectories
 that are geodesics on the moduli space.  The
trajectories are described by the collective coordinates of the moduli
space corresponding to the parameters of a monopole
solution that can be varied without altering the
topological charge where gauge transformed solutions
are identified.  Equivalently, zero modes satisfy the
equations of motion with the potential energy
at its minimum, and their kinetic energy describes
this geodesic motion.  By limiting the dynamics to zero modes,
the particles of the model (photons, massive gauge bosons,
 scalar fields, fermions) and internal massive excitations
of the monopole are excluded from the picture.

The non-trivial monopole solution louses up two of the
supersymmetries which are transformed into fermion zero
modes.  These two supersymmetries correspond to
 the collective coordinates of one monopole, and one can further
construct two commuting fermion zero modes for each of the other
collective coordinates.  The fermion modes are constructed
from bosonic zero modes and eight orthonormal spinors
 such that the number of
independent bosonic
modes is half that of the fermionic ones. The commuting
fermion modes are paired with collective grassmann variables
whose time dependence represents the motions of
the fermion zero modes on the moduli space.
  Since two of the four
supersymmetries leave the monopole solution invariant,
the number of unblemished supersymmetries is that
needed for the description of trajectories on the
moduli space to be an $N=4$ quantum mechanics.

\newsec{The Monopole Solution of N=4 Yang-Mills Theory}

In this section we discuss the N=4 Yang-Mills theory in three
 space plus one time dimensions and its monopole solution to
the equations of motion.  The notations used with some minor changes
are from Osborn \fiv.  The Lagrangian for this theory is
\eqn\one{\eqalign{{\cal L}&= ({1\over 4} F_{\mu\nu}F^{\mu\nu}-{1\over 2}
D_{\mu}{\cal A}_i D^{\mu}{\cal A}_i\cr &
-{1\over 2} D_{\mu}{\cal B}_{\dot j} D^{\mu}{\cal B}_{\dot j}+
{1\over 2} i\bar \Psi\gamma^{\mu}D_{\mu}\Psi\cr &+{1\over 2}
\bar\Psi[\alpha^i {\cal A}_i+i\gamma_5\beta^{\dot j}{\cal B}_{\dot j},\Psi])
-V({\cal A},{\cal B})\cr}}
where the indices $i,\dot j =1,2,3$, and
\eqn\two{\eqalign{V({\cal A},{\cal B})&={-1\over 4}
\left([{\cal A}_i,{\cal A}_j][{\cal A}_i,{\cal A}_j]
+[{\cal B}_{\dot i},{\cal B}_{\dot j}][{\cal B}_{\dot i},{\cal B}_{\dot j}]+
2[{\cal A}_i,{\cal B}_{\dot j}][{\cal A}_i,{\cal B}_{\dot j}]\right)\cr
F_{\mu\nu}&=\partial_{\mu}A_{\nu}-\partial_{\nu}A_{\mu}+[A_{\mu},
A_{\nu}]\cr}} with $ D_{\mu}=\partial_{\mu}+[A_{\mu},\; \;]$.
  All fields are in the adjoint representation of $SU(2)$
 (e.g. $A_{\mu}=A_{\mu}^A T^A$), and $(T^A T^B)={1\over 2}
{\rm tr}(T^A T^B)= -\delta^{AB}$.  The metric used
is $g_{ij}=-\delta_{ij}$ for spatial indices
while $g_{00}=1$.
  The $4\times 4$ matrices $\alpha ^i,\beta ^{\dot j}$ satisfy
the following relations:
\eqn\three{\eqalign{S^{ij}=[\alpha^i,\alpha^j]&=-2\epsilon^{ijk}\alpha^k\cr
V^{\dot i\dot j}=[\beta^{\dot i},\beta^{\dot j}]&=
-2\epsilon^{\dot i\dot j\dot k}\beta^{\dot k}\cr
U^{i\dot j}=\lbrace\alpha^i ,\beta^{\dot j}\rbrace&=-U^{\dot j i}\cr
[\alpha^i,\beta^{\dot j}]&=0\cr
\lbrace\alpha^i,\alpha^j\rbrace&=-2\delta^{ij}\cr
\lbrace\beta^{\dot i},\beta^{\dot j}\rbrace&
=-2\delta^{\dot i\dot j}.\cr}}
The fermions $\Psi_{tu}$ are Majorana spinors with $t$ a Lorentz
spinor index acted on by the $\gamma^{\mu}$ and $u$ an $SU(4)$
index transforming
under $S^{ij}$, $V^{\dot i\dot j}$, and $iU^{i\dot j}$
such that
$\bar\Psi=\Psi^T C=\Psi^{\dagger}\gamma^0$
where $C$ is the charge conjugate matrix.  The above Lagrangian
can be derived from the ten dimensional supersymmetric
Yang-Mills action, and a global $SO(6)\sim SU(4)$
symmetry exists as a consequence of the reduction
 of the Lorentz group \sox :
\eqn\glo{\eqalign{\delta A_{\mu}&=0\cr
\delta\Psi&={-1\over 8}[S^{ij}\epsilon^S_{ij}+V^{\dot i\dot j}
\epsilon^V_{\dot i\dot j}+i\gamma_5 U^{i\dot j}\epsilon^U_{i\dot j}
]\Psi\cr
\delta{\cal A}_i&=\epsilon^S_{ij}{\cal A}_j+\epsilon^U_{i\dot j}
{\cal B}_{\dot j}\cr
\delta{\cal B}_{\dot i}&=\epsilon^U_{\dot i j}{\cal A}_j+
\epsilon^V_{\dot i\dot j}{\cal B}_{\dot j}\cr}}
where $\epsilon^S_{ij}$, $\epsilon^V_{\dot i\dot j}$, and
$\epsilon^U_{i\dot j}$ are constant and antisymmetric in
their indices.
 The action is also invariant
under the following $N=4$ supersymmetries
\eqn\four{\eqalign{\delta\Psi&=[{i\over 2}\gamma^{\mu\nu}
F_{\mu\nu}+\gamma^{\mu}D_{\mu}(\alpha^i {\cal A}_i
+i\gamma_5\beta^{\dot j} {\cal B}_{\dot j})
\cr&-{1\over 2}i\epsilon^{ijk}\alpha^k [{\cal A}_i,{\cal A}_j]
-{1\over 2}i
\epsilon^{\dot i\dot j\dot k}\beta^{\dot k} [{\cal B}_{\dot i},
{\cal B}_{\dot j}]\cr&-\alpha^i\beta^{\dot j}
[{\cal A}_i,{\cal B}_{\dot j}]\gamma_5] \epsilon\cr \delta A_{\mu}&=
\bar\epsilon\gamma_{\mu}\Psi\cr\delta {\cal A}_i&=-i\bar
\epsilon\alpha^i\Psi\cr\delta {\cal B}_{\dot j}&=
\bar\epsilon\beta^{\dot j}\gamma_5\Psi
\cr}}
  where $\epsilon_{tu}$
is a constant, anticommuting Majorana spinor.

The equation of motion for the gauge field is
\eqn\motion{D^{\nu}F_{\mu\nu}-[{\cal A}_i,D_{\mu}{\cal A}_i]
-[{\cal B}_{\dot j},D_{\mu}{\cal B}_{\dot j}]
-{1\over 2}i[\bar\Psi,\gamma_{\mu}\Psi]=0.}
Following
the methods of Harvey, Strominger \uno   and Gauntlett \duo  ,
we want to expand
solutions of this equation in $n=n_{\partial_0}+{n_f\over 2}$ where
$n_{\partial _0}$ is the number of time derivatives and $n_f$
 is the number of fermion fields.  To zeroth order in $n$ the
solution as discussed by various authors \sev  is a static monopole:
\eqn\six{\eqalign{&B_i=D_i\Phi\cr &A_0=0\cr &V({\cal A},{\cal B})=0\cr
&(\Phi^2)=v^2|_{| \buildrel \rightharpoonup \over x|
\rightarrow \infty}\cr}}
where the scalar field $\Phi$ is defined as a function
of the ${\cal A}_i$,${\cal B}_{\dot j}$
\eqn\seven{\eqalign{&\Phi=a_i{\cal A}_i+b_{\dot j}{\cal B}_{\dot j}\cr
&a_ia_i+b_{\dot j}b_{\dot j}=1,\cr}}
$a_i,b_{\dot j}$ constant, and the nonabelian magnetic field $B_i$ is
\eqn\eight{B_i={1\over 2}\epsilon_{ijk}F_{jk}.}
The scalar field equations also are satisfied since
$D_iB_i=0$.
Introducing Pauli matrices $\sigma _i$ and a
projection matrix $P$,
\eqn\nine{\eqalign{&\sigma _i={i\over 2}\epsilon_{ijk}
\gamma^{jk}=-\gamma ^i\gamma ^0\gamma _5\cr
& P=P^{\dagger}=\gamma ^0\gamma _5\alpha ^i a_i+
i\gamma ^0\beta ^{\dot j} b_{\dot j}\cr}}
we see that
\eqn\ten{\delta\Psi=\sigma _i B_i(1+P)\epsilon}
and that this solution has broken half (those having $P\epsilon
=\epsilon$)
of the supersymmetries.  The solution has also reduced the
gauge symmetry from $SU(2)$ to $U(1)$ by specifying a
direction in isospin space for the monopole on the two-sphere
spatial boundary.  The global symmetry is reduced to $SO(5)$ by
this solution.  To zeroth order in $n$ the Lagrangian
is a topological charge invariant \eig  under smooth deformations
of $\Phi$, $A_i$ since
\eqn\eleven{\eqalign{L^{(0)}&=\int _{R^3} d(F\Phi)
\cr &=-v\int _{S^2(\infty)}({1\over v}F \Phi
-{1\over 2v^3}\Phi D\Phi\wedge D\Phi)=-4\pi kv\cr}}
 where $k$ is an integer,
 and the second term does
not contribute because $D\Phi\sim O({1\over r^2})$
  for smooth, finite energy configurations.  The
absolute value of this Lagrangian
is the rest energy (mass) of $|k|$  monopoles or antimonopoles
 (for $k$ negative).

\newsec{Bosonic Zero Modes and Geometry on the Monopole
Moduli Space}

The goal is to determine the low energy dynamics of monopoles.
To accomplish it we need to consider bosonic zero modes and the
geometry of the monopole moduli space \nin .  For a monopole solution
in $R^3$ with topological number $k$, there are $4k$ independent
directions in which the solution can be deformed continuously,
thus, preserving
its topology or monopole number.  These parameters correspond for $k=1$
to the position of the monopole in $R^3$ and the parameter of
the unbroken $U(1)$ gauge symmetry for gauge transformations
with noncompact support.  For general $k$ the monopole moduli
manifold factorizes as $M^k=R^3\times{S^1\times M^k_0\over Z_k}$
such that the center of mass sits in $R^3$, and
 the $S^1$ angle plays the same role as for $k=1$.  The $4k-4$
dimensional manifold $M^k_0$ is coordinatized by parameters governing
the relative motion of $k$ monopoles.  Only the $M^2_0$
metric is known.  The $Z_k$ identification reflects the fact
that the $k$ monopoles are indistinguishable.  For large distances
$M^k$ approximates $k$ copies of $M^1$ physically representing
$k$ noninteracting monopoles.  The manifolds $M^k$ and $M^k_0$
are hyperkahler implying that there exists a triplet of
covariantly constant complex structures with the algebra
\eqn\one{J_i J_j=-\delta _{ij}-\epsilon _{ijk}J_k.}

We introduce time dependence by allowing the $4k$ parameters
to become collective coordinates that depend on time.  The time
derivative of the $U(1)$ parameter is proportional to the total
electric charge of the monopoles.  Electrically charged monopoles
are called dyons.
  The fields which depend on these collective coordinates
also gain a time dependence.  Consider the $k$-monopole sector
with $4k$ collective coordinates $X^{\alpha}(t)$.  The fields
$A_i(x,X^{\alpha})$ and $\Phi(x,X^{\alpha})$ depend on these as
well as on $R^3$.  There are, thus, bosonic zero modes $\delta_{\alpha}
A_i$,
$\delta_{\alpha}\Phi$ tangent to the moduli manifold.  They are
defined as follows:
\eqn\two{\eqalign{&\delta _{\alpha} A_i=[s_{\alpha},D_i]\cr
&\delta _{\alpha}\Phi=[s_{\alpha},\Phi]\cr}}
where $s_{\alpha}=\partial _{\alpha}+[\epsilon _
{\alpha},\;\;]$.
The gauge parameter $\epsilon_{\alpha}$ is fixed by requiring
\eqn\gauge{D^i\delta _{\alpha}A_i-[\Phi,\delta _{\alpha}
\Phi]=0,}
 and equation \motion\  is still satisfied.  To lie in the tangent
space these zero modes must also satisfy
\eqn\zmodes{\epsilon _{ijk}D^j\delta _{\alpha}A^k=
D_i\delta _{\alpha}\Phi +[\delta _{\alpha}A_i,\Phi].}

The various geometrical structures on the moduli space are induced
from these bosonic zero modes.  The metric is a symmetric, nondegenerate
 tensor product of covectors.  Since the zero mode equations
\gauge\ , \zmodes\ transform covariantly with respect to
diffeomorphisms of $R^3$, global rotations of the gauge group,
 and global rotations of $(D_i,\Phi)\equiv (D_m)$, the metric
should be invariant under these metamorphoses.  The simplest
Riemannian metric satisfying these conditions is
\eqn\metric{g_{\alpha\beta}=-\int d^3 x\,(\delta _{\alpha}
A_i\delta _{\beta}A_i+\delta _{\alpha}\Phi\delta _{\beta}\Phi).}
  The three complex structures are $J_{i\alpha}\,^{\beta}$ where
\eqn\J{J_{i\alpha\beta}=2\int d^3 x\,(\delta _{[\alpha
}A_i\delta _{\beta]}\Phi-{\epsilon _{ijk}\over 2}
\delta _{[\alpha}A_j\delta _{\beta]}A_k)}
  and $i=1,2,3$.  In addition
to zero modes there are non-zero modes which together satisfy
a completion condition
\eqn\compl{\delta ^{\alpha}{C_m^A (x)}^p\delta _{\alpha}{C_n^B (y)}^p=
\delta _{mn} \delta^{AB} \delta (x-y)}
  where $C_m=A_i,\Phi$; $m=1,2,3,4$,
 and $p$ indexes the modes.
  The Christoffel symbols and Riemann
curvature tensor can be calculated to take the form
\eqn\christo{\Gamma _{\alpha\beta\gamma}=
-\int d^3 x\,(\delta _{\alpha} C_m [s_{\gamma},\delta
_{\beta}C_m ])}

\eqn\riemm{R_{\alpha\beta\gamma\delta}=\int d^3 x\,(
\delta _{\alpha} C_m [\Phi _{\gamma\delta},\delta _{\beta} C_m ])}
  where $\Phi _{\gamma\delta}=[s_{\gamma},s_{\delta}]$.
In these calculations we must remember that zero
modes are orthogonal to nonzero modes and that only zero
modes make up tensors that are tangent to the moduli manifold.
Using \gauge\ , \zmodes\ , and \compl\ we can show
that the complex structures are
covariantly constant and follow the correct algebra.
We have started with $4k\times 4$ bosonic zero modes, but
using the complex structures to relate the modes
through \compl\  shows that only $4k$ of them are independent.

\newsec{Fermion Zero Modes and the Low Energy Lagrangian of $N=4$
Supersymmetric Monopoles}

Let us now extend the picture to fermions and write down the
supersymmetric quantum mechanics of $N=4$ monopoles.  Half of the
supersymmetries are destroyed by the monopole solution, and these eight
destroyed supersymmetries satisfy a zero mode equation
\eqn\zmf{\gamma ^i D_i\delta\Psi=i\gamma _5\gamma ^0 P
[\Phi,\delta\Psi].}
The Majorana condition $B\Psi ^* =\Psi$
where $B=\gamma ^0 C^{-1}$, $B{(\gamma ^{\mu})}^*=-\gamma
 ^{\mu}B$, and $PB=BP^*$ cuts the number of supersymmetries
 in half.  As in the bosonic case there are fermionic fluctuations
that satisfy the zero mode equation.  These modes can be
written  explicitly in complex coordinates as
\eqn\zmftwo{\Psi_{\beta}^{as}=(\sigma _i \delta _{\beta}A_i
-i\delta _{\beta}\Phi)({1+\sigma _2 \over 2})\epsilon ^{as}}
  and
\eqn\zmfthree{\Psi _{\beta^ *}^{as\star}=B(\Psi _{\beta}^{as})^*}
where
 $\epsilon_{tu}^{as}$ is a commuting spinor.  The
broken supersymmetries are a linear combination
of the eight $R^3\times S^1$ modes.  The precise
form of these solutions has been chosen so that they are
eigenfunctions of the complex structure $J_2$:
\eqn\Jtwo{\eqalign{&J_{2\alpha}\, ^{\beta}\Psi _{\beta}^{as}=i
\Psi _{\alpha}^{as}\cr &J_{2\alpha^ *}\, ^{\beta^ *}
\Psi _{\beta^ *}^{as\star}=-i
\Psi _{\alpha^ *}^{as\star}.\cr}}
We are using kahlerity to set $g_{\alpha\beta}=J_{2\alpha\beta}=0$.
  The indices $a,s$ are
both two dimensional.  In fact, since $[P,\sigma_i]=0$ we let
\eqn\pauli{\sigma _i\epsilon _{as}=\epsilon _{as'}(\sigma _i)_{s's}}
  with
\eqn\eps{\eqalign{&\epsilon ^{\dagger}_{as}\epsilon _{a's'}=
\delta _{aa'}\delta _{ss'}\cr
&\epsilon ^{\dagger}_{as}B\epsilon _{a's'}^*=0\cr
& P\epsilon _{as}=\epsilon _{as}.\cr}}
For the purposes of dimensional reduction we introduce
matrices $\rho_i$ dependent on the choice of
$a_i$, $b_{\dot j}$ \seven\ so that the gamma matrices
take the following form:
\eqn\ga{\eqalign{&\gamma_i=\gamma_5\gamma_0\sigma_i\cr &
\gamma_0=A\rho_2\cr}}
where $\lbrace A,\rho_i\rbrace=0$, $[\rho_i,\sigma_j]=0$, $A^2=-1$, and
the $\rho_i$ act on the index $a$ analogously to \pauli\ .
   In these coordinates the fermion zero modes $\Psi_{\beta^ *}
^{as}$ and their charge conjugate modes are both zero.
Also, the modes obtained by reversing the projection
 ($\sigma_2\rightarrow -\sigma_2$) are either zero
or can be obtained from the listed ones by
multiplication of the matrix $J_3$.
The number of zero modes agrees with the Callias index
theorem \dix
 because taking into account the projection ${1+\sigma_2}\over 2$
and the Majorana condition \zmfthree\ , we are left with $2k\times 4$
zero modes.  As mentioned above eight of the modes corresponding
to the $R^3\times S^1$ coordinates come from the broken supersymmetries.

The fermion zero modes do not alter the vacuum energy or mass
of the monopoles, and we introduce time dependence analogously
to the bosonic case by pairing the zero modes with anticommuting
collective parameters $\lambda^{\alpha}_{as}(t)$.  Then,
\eqn\psi{\Psi=\Psi _{\alpha}^{as}\lambda _{as}^{\alpha}
+\Psi _{\alpha^ *}^{as\star}\lambda ^{\alpha^ * T}_{as}}
   and $J_{2\alpha}\, ^{\beta}\lambda ^{\alpha}_{as}
=i\lambda ^{\beta}_{as}$.

The next step is to solve \motion\ to first order in time ($n=1$).
The following equation which can be derived using $J_2$ is helpful:
\eqn\help{\eqalign{\lbrace 2i[\delta_{(\alpha^*}A_2 ,\delta_
{\beta )}\Phi]&-i\epsilon_{2ij}[\delta_{(\alpha^*}A_i ,\delta_
{\beta)}A_j]\rbrace \bar\lambda^{\alpha^*}\gamma_0(1+\sigma_2)
\lambda^{\beta}=\cr
\lbrace [\delta_{\alpha^*}A_i,\delta_{\beta}A_i]&+
[\delta_{\alpha^*}\Phi,\delta_{\beta}\Phi]\rbrace
\bar\lambda^{\alpha^*}\gamma_0(1+\sigma_2)
\lambda^{\beta}.\cr}}
The solution after some algebra is
\eqn\nequalone{\eqalign{&A_0=\epsilon_{\alpha}\partial_0 Z^{\alpha}+
\epsilon_{\alpha^ *}\partial_0 Z^{\alpha^ *}+
{i\over 2}\Phi_{\alpha^ *\beta}\bar\lambda^{\alpha^ *}
\gamma_0(1+\sigma_2)\lambda^{\beta}\cr
&F_{0i}=\partial_0 Z^{\alpha}\delta_{\alpha}A_i+
\partial_0 Z^{\alpha^ *}\delta_{\alpha^ *}A_i+
is_{[\alpha^ *}\delta_{\beta]}A_i\bar\lambda
^{\alpha^ *}\gamma_0(1+\sigma_2)\lambda^{\beta}\cr}}
   where $\gamma^0=\rho_2$
acting on the $a$ index.  The fermionic term of $F_{0i}$ is
not a zero mode and will not contribute to the low energy
dynamics.

We are ready to expand the Lagrangian to order $n=2$.  Recalling
that the moduli space is a Kahler manifold; using \metric\ ,
\J\ , \christo\ , \riemm\ , \help\ ;
and substituting
\psi\ , \nequalone\ in the kinetic energy; the result is
\eqn\lag{\eqalign{L^{(2)}&=\partial_0 Z^{\alpha^ *}
\partial_0 Z^{\beta}g_{\alpha^ *\beta}-{i\over 2}
(\bar\lambda^{\alpha^ *}\gamma^0 (1+\sigma_2)D_0 \lambda^{\beta}
g_{\alpha^ *\beta}+c.c.)\cr &-{1\over 2}\bar\lambda^{\alpha^ *}
\gamma^0 (1+\sigma_2)\lambda^{\beta}\bar\lambda^{\gamma^ *}
\gamma^0 (1+\sigma _2)\lambda^{\delta}R_{\alpha^ *\beta
\gamma^ *\delta}\cr}}
 where
\eqn\D{D_0\lambda^{\alpha}=\partial_0\lambda^{\alpha}+
\Gamma^{\alpha}_{\beta\gamma}\partial_0 Z^{\beta}
\lambda^{\gamma}.}
  Substituting
\eqn\chvar{\lambda^{\beta}_{as}={1\over 2}{-i\choose 1}
\lambda_a^{\beta}+{1\over 2}{i\choose 1}\lambda_a^{\beta\, -}}
  and writing $L^{(2)}$ in terms of
real coordinates yields
\eqn\lagtwo{L^{(2)}={1\over 2}g_{\alpha\beta}\partial_0
X^{\alpha}\partial_0 X^{\beta}-{i\over 2}\bar\lambda^{\alpha}
\gamma^0 D_0\lambda^{\beta}g_{\alpha\beta}-{1\over 12}
R_{\alpha\beta\gamma\delta}\bar\lambda^{\alpha}\lambda
^{\gamma}\bar\lambda^{\beta}\lambda^{\delta}}
   with a Fierz rearrangement
of the curvature term.  This Lagrangian represents a quantum
mechanical system with $N=4$ supersymmetry.  The supersymmetry
transformations that leave the action invariant are:
\eqn\suptrf{\eqalign{&\delta X^{\alpha}=J_{\mu\,\,\beta} ^{\,\alpha}
\bar\epsilon^{\mu}\lambda^{\alpha}\cr &\delta\lambda^{\alpha}=
i{(J_{\mu\,\,\beta} ^{\,\alpha})}^{-1}\gamma^0\partial_0
X^{\beta}\epsilon^{\mu}-\Gamma^{\alpha}_{\beta\gamma}\delta X^
{\beta}\lambda^{\gamma}\cr}}
where $J_{0\,\,\beta}^{\alpha}=\delta^{\alpha}
_{\,\beta}$, $\mu=0,1,2,3$ and $\epsilon^{\mu}$
are real, two-dimensional anticommuting
spinors.

\newsec{Conclusions}

After quantization the energy of a dyon can be written as
\eqn\energy{E=4\pi v+2\pi vu^2+{{\hbar}^2 n^2\over 8\pi}v}
where $u$ is the velocity of the dyon, and $q_e=\hbar n$
is the electric charge of the dyon.  Since the charge
$q_e$ is conserved, dyons have a mass
\eqn\mass{M_{dyon}=M_{monopole}+{\hbar^2 n^2\over 8\pi}v}
The energy is, of course, nonrelativistic, and one would expect
relativistically $E=\sqrt{p^2+M^2_{monopole}}$ since the
excess mass of the dyon is part of the kinetic energy.
  Taking $u=0$ implies relativistically that
\eqn\mrel{M^{rel}_{dyon}=4\pi v\sqrt{1+{\hbar^2 n^2\over
{(4\pi)}^2}}=v\sqrt{q_m^2+q_e^2}}
where the magnetic charge $q_m=4\pi$ in these units.  This is
indeed the formula derived from consideration of the central charges
of the $N=4$ supersymmetry \fiv .

One motivation for finding this low energy action of
monopoles is that their quantum scattering can then
be calculated.  Gibbons and Manton \thr have performed these
computations for the bosonic two-monopole.  Some effort has
been applied to extend their results to the
supersymmetric case.  By comparing these results to low energy
scattering in the particle sector of the theory,
 one can possibly find evidence for the duality
conjecture of Montonen and Olive \ele .  This conjecture
postulates an exchange of monopole and particle dynamics
 under the interchange of electric and magnetic
charges and the inversion of the coupling constant
$g\rightarrow {1\over g}$.  Here,  we have taken
$g=4\pi$.  The relativistic mass formula is valid
for all particles and solitons.  Osborn \fiv  has shown that
the monopole supermultiplet contains the same spins
as the particle one.  Further endeavors to search for
evidence of duality are being undertaken.

\bigskip\centerline{\bf Acknowledgements}\nobreak

My appreciation goes to Jeff Harvey for suggesting this
problem and for conversations.  This work was supported
in part by NSF contract PHY91-23780.

\listrefs
\end